\documentclass[aps,prl,twocolumn,superscriptaddress,longbibliography]{revtex4-2}
\usepackage{amsmath,amssymb}
\usepackage[pdftex]{hyperref,graphicx}
\hypersetup{colorlinks = true, urlcolor = blue, linkcolor = blue, citecolor = blue}
\usepackage{xcolor}
\usepackage{bm}
\usepackage {verbatim}

\usepackage[etex=true,export]{adjustbox}
\usepackage{makecell}
\usepackage{multirow}
\usepackage{tabularx,multirow,array,diagbox}
\usepackage{adjustbox}
\usepackage{booktabs}

\newcommand{\beq}{\begin{equation}}
\newcommand{\eeq}{\end{equation}}
\newcommand{\beql}{\begin{equation*}}
\newcommand{\eeql}{\end{equation*}}
\newcommand{\beqn}{\begin{eqnarray}}
\newcommand{\eeqn}{\end{eqnarray}}

\renewcommand{\vec}[1]{\mbox{\boldmath$#1$}}

\begin{document}
\title{Majorana Zero Modes in Twisted Transition Metal Dichalcogenide Homobilayers}
\author{Xun-Jiang Luo}
\affiliation{School of Physics and Technology, Wuhan University, Wuhan 430072, China}
\author{Wen-Xuan Qiu}
\affiliation{School of Physics and Technology, Wuhan University, Wuhan 430072, China}
\author{Fengcheng Wu}
\email{wufcheng@whu.edu.cn}
\affiliation{School of Physics and Technology, Wuhan University, Wuhan 430072, China}
\affiliation{Wuhan Institute of Quantum Technology, Wuhan 430206, China}

\begin{abstract}
Semiconductor moir\'e superlattices provide a highly tunable platform to study the interplay between electron correlation and band topology. For example, the generalized Kane-Mele-Hubbard model can be simulated by the topological moir\'e flat bands in twisted transition metal dichalcogenide homobilayers. In this system, we obtain the filling factor, twist angle, and electric field-dependent quantum phase diagrams with a plethora of phases, including the quantum spin Hall insulator, the in-plane antiferromagnetic state, the out-of-plane antiferromagnetic Chern insulator, the spin-polarized Chern insulator, the in-plane ferromagnetic state, and the 120$^\circ$ antiferromagnetic state. We predict that a gate-defined junction formed between the quantum spin Hall insulator phase with proximitized superconductivity and the magnetic phases with in-plane magnetization (either ferromagnetism or antiferromagnetism) can realize one-dimensional topological superconductor with Majorana zero modes. 
Our proposal introduces semiconductor moir\'e homobilayers as an electrically tunable Majorana platform with no need of an external magnetic field.
\end{abstract}
\maketitle

\textit{Introduction}.---
Transition metal dichalcogenide (TMD) moir\'e systems have attracted great research interest \cite{Wu2018,Wu2019,Naik2018,Wang2020,Tang2020,Regan2020,Xu2020,Li2021a, Li2021,Zhang2022,Zhao2023,Pan2020,Pan2020a,Devakul2021,Zhang2021,Pan2022,Devakul2022,Xie2022a,Chang2022,Dong2023,Zhao2023,Wu2023,Xie2023,Luo2023}, as they provide a highly tunable platform for studying electron correlation, band topology, as well as their interplay.  Mott insulators, generalized Wigner crystals, and quantum anomalous Hall insulators were experimentally observed in TMD moir\'e heterobilayers \cite{Wang2020,Tang2020,Regan2020,Xu2020,Li2021a,Li2021,Zhang2022,Zhao2023}. Twisted TMD homobilayers were predicted to host intrinsic topological moir\'e bands that can simulate the generalized Kane-Mele model \cite{Wu2019}.  Recent optical and transport experiments reported observation of not only integer but also fractional quantum anomalous Hall insulators in twisted bilayer MoTe$_2$  ($t$MoTe$_2$) \cite{Anderson2023, Cai20231,Zeng20231,Park20231,Xu20231}.   Signatures of the integer quantum anomalous Hall insulator were also detected in twisted bilayer WSe$_2$ ($t$WSe$_2$) using scanning single electron transistor microscopy \cite{Foutty2023}. These exciting experimental findings immediately stimulated extensive theoretical investigations \cite{Wang2023,Reddy20231,Qiu20231,Dong20231,Goldman20231,NM2023}.

\begin{figure}
\centering
\includegraphics[width=3.5in]{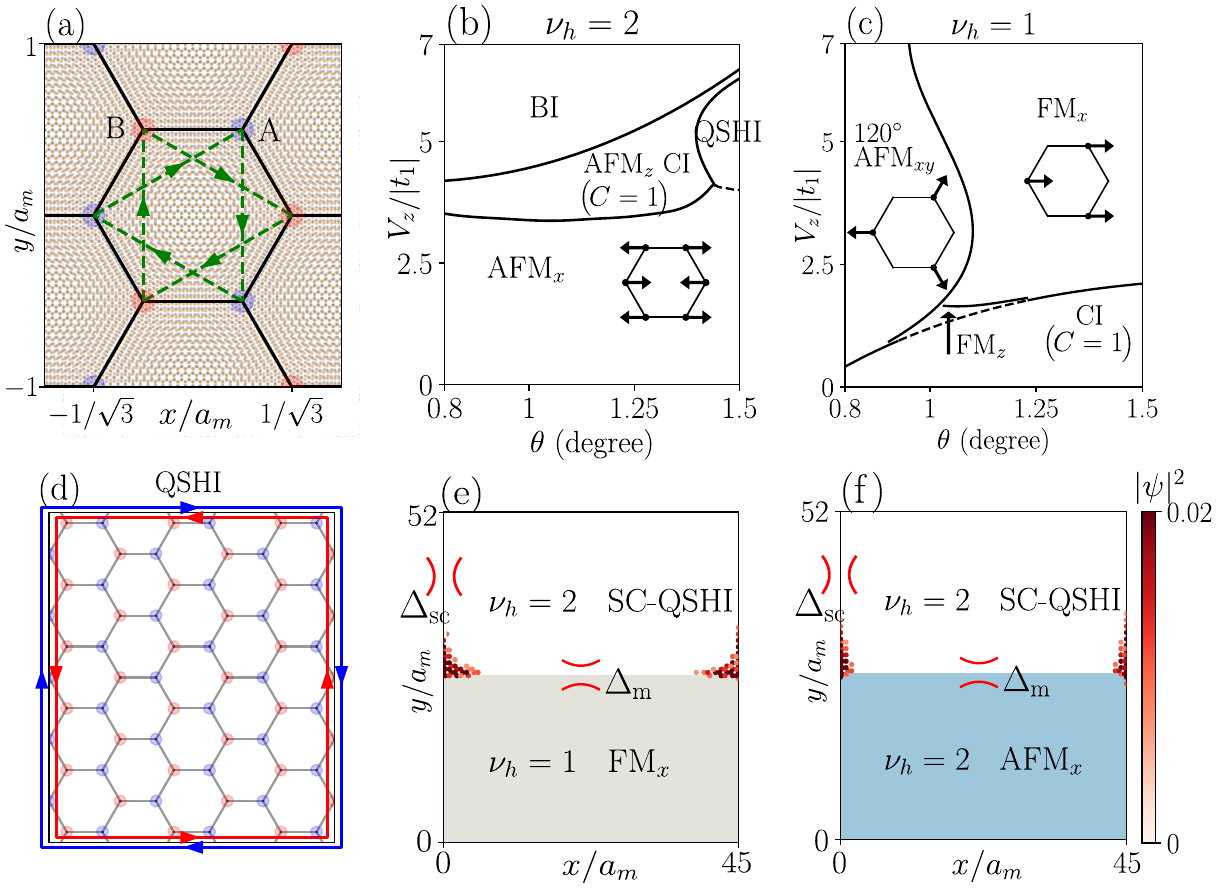}
\caption{(a) Illustration of the generalized Kane-Mele model on the effective honeycomb lattice formed in twisted TMD homobilayers.  $a_m$ is the  moir\'e lattice constant. The green arrows denote the complex next nearest neighbor hopping. (b), (c) Quantum phase diagrams at $\nu_h=2$ and $\nu_h=1$, respectively.  The solid (dashed) black lines mark the first-order (continuous) phase transition. The arrows on the honeycomb lattices represent the spin configuration. (d) Schematic illustration of the QSHI state.  (e), (f) Spatial distribution of two MZMs in SC-QSHI/AFM$_x$ and SC-QSHI/FM$_x$ junctions, respectively. In (e) and (f), model parameters are fixed at the green points in Fig.~\ref{fig2}(b) and Fig.~\ref{fig4}(b), respectively.}
\label{fig1}
\end{figure}

An appealing feature of twisted TMD homobilayers is that they host rich quantum phase diagrams that are experimentally tunable 
\cite{Anderson2023,Cai20231,Zeng20231,Park20231,Xu20231,Foutty2023}.  This is illustrated in Figs.~\ref{fig1}(b) and \ref{fig1}(c), which show, respectively, our theoretical phase diagrams of $t$WSe$_2$ at hole filling factors $\nu_h=2$ and $\nu_h=1$  as functions of twist angle $\theta$ and vertical electric field-induced potential $V_z$. Here the results are obtained by a mean-field study of a generalized Kane-Mele-Hubbard model. At $\nu_h=2$, the theoretical phase diagram includes the quantum spin Hall insulator (QSHI), the in-plane antiferromagnetic state (AFM$_x$), the out-of-plane antiferromagnetic Chern insulator (AFM$_z$ CI) with a Chern number $C$ of 1, and the band insulator (BI). At $\nu_h=1$, the theoretical phase diagram hosts the spin-polarized Chern insulator (CI) with $C=1$, the topologically trivial out-of-plane ferromagnetic state (FM$_z$), the in-plane ferromagnetic state (FM$_x$), and the 120$^{\circ}$ antiferromagnetic state (120$^{\circ}$ AFM$_{xy}$). The phase transition between CI and topologically trivial magnetic insulators driven by $V_z$ has recently been experimentally realized in both $t$WSe$_2$ and $t$MoTe$_2$ at $\nu_h=1$ \cite{Anderson2023,Cai20231,Zeng20231,Park20231,Xu20231,Foutty2023}. Since multiple phases can be realized within one system of a given $\theta$ by tuning $\nu_h$ and $V_z$, it is a promising direction to explore new physics that can emerge in junctions formed between different phases. 

In this Letter, we theoretically predict that a planar junction formed between the QSHI phase with proximitized superconductivity (SC-QSHI) and the magnetic phases with in-plane magnetization (e.g., FM$_x$ and AFM$_x$ phases) can realize one-dimensional topological superconductor (TSC) with Majorana zero modes (MZMs) at the boundaries, as illustrated in Figs.~\ref{fig1}(e) and \ref{fig1}(f). The MZMs are highly sought after due to their non-Abelian braiding statistics and potential application in topological quantum computation \cite{Nayak2008}. Although MZMs are actively studied in several condensed matter
platforms such as semiconductor nanowires \cite{Sau2010,Lutchyn2010,Oreg2010,Mourik2012,Alicea_2012}, superconducting vortex cores in topological insulators \cite{Fu2008,Jin2016,Xu2016,Wang2018,Mercado2022}, and ferromagnetic atomic chains \cite{Nadj-Perge2014,Jeon2017}, the deterministic evidence of MZMs is still an ongoing research topic. Therefore, it is worthwhile to search for new platforms to realize MZMs. Our theoretical proposal introduces TMD moir\'e systems as a potential Majorana platform based on two-dimensional materials \cite{San-Jose2015, Thomson2022,Xie2023a}, where MZMs can be tuned electrically without the need of any external magnetic field.

\textit{Model Hamiltonian and phase diagrams}.---
We start with the generalized Kane-Mele-Hubbard model that can phenomenologically capture the interacting physics in topological bands of twisted TMD homobilayers \cite{Wu2019}, 
\beqn
H=H_{\text{KM}}+U\sum_{i,\alpha}\hat{n}_{i\alpha\uparrow}\hat{n}_{i\alpha\downarrow}+\frac{V_z}{2}\sum_{i,\alpha}\ell_{\alpha}\hat{n}_{i\alpha}.
\label{Hm}
\eeqn
Here $H_{\text{KM}}$ is the generalized Kane-Mele Hamiltonian,
\beqn
&&H_{\text{KM}}= t_1 \sum_{\langle ij\rangle,\alpha\neq \beta,s} c^\dagger_{i\alpha s} c_{j\beta s} + \nonumber\\
&&\quad\quad\quad |t_2| \sum_{\langle \langle ij\rangle\rangle,\alpha,s} e^{i\phi s \epsilon_{ij}} c^\dagger_{i\alpha s} c_{j\alpha s}
+\cdots,
\eeqn
where subscripts $\alpha$ and $\beta$ denote the A or B sublattice in the honeycomb lattice [Fig.~\ref{fig1}(a)], $s$ is the spin index that is locked to the valley index in TMDs, $t_1$ ($t_2$) is the hopping parameter between the nearest (next-nearest) neighbor sites, and the symbol $\cdots$ contains longer-range hopping terms. The next-nearest neighbor hopping parameters carry spin- and sublattice-dependent  phase factors  $e^{i\phi s \epsilon_{ij}}$, where $s=+1$ $(-1)$ for spin up (down) and $\epsilon_{ij}=\pm 1$ for different hopping paths [Fig.~\ref{fig1}(a)]. In Eq.~\eqref{Hm}, $U$ represents the onsite Hubbard repulsion, while the last term is a staggered sublattice potential $\ell_\alpha V_z/2$ with $\ell_{\text{A}}$= $1$ and $\ell_{\text{B}}$= $-1$. This staggered potential is generated by an applied vertical electric field since the two sublattices are polarized to opposite layers \cite{Wu2019}. The hopping parameters are tuned by the twist angle $\theta$. Throughout this paper, we use the hopping parameters derived for $t$WSe$_2$ \cite{Devakul2021,supp}, while the obtained results are also applicable to $t$MoTe$_2$.

\begin{figure}
\centering
\includegraphics[width=3.5in]{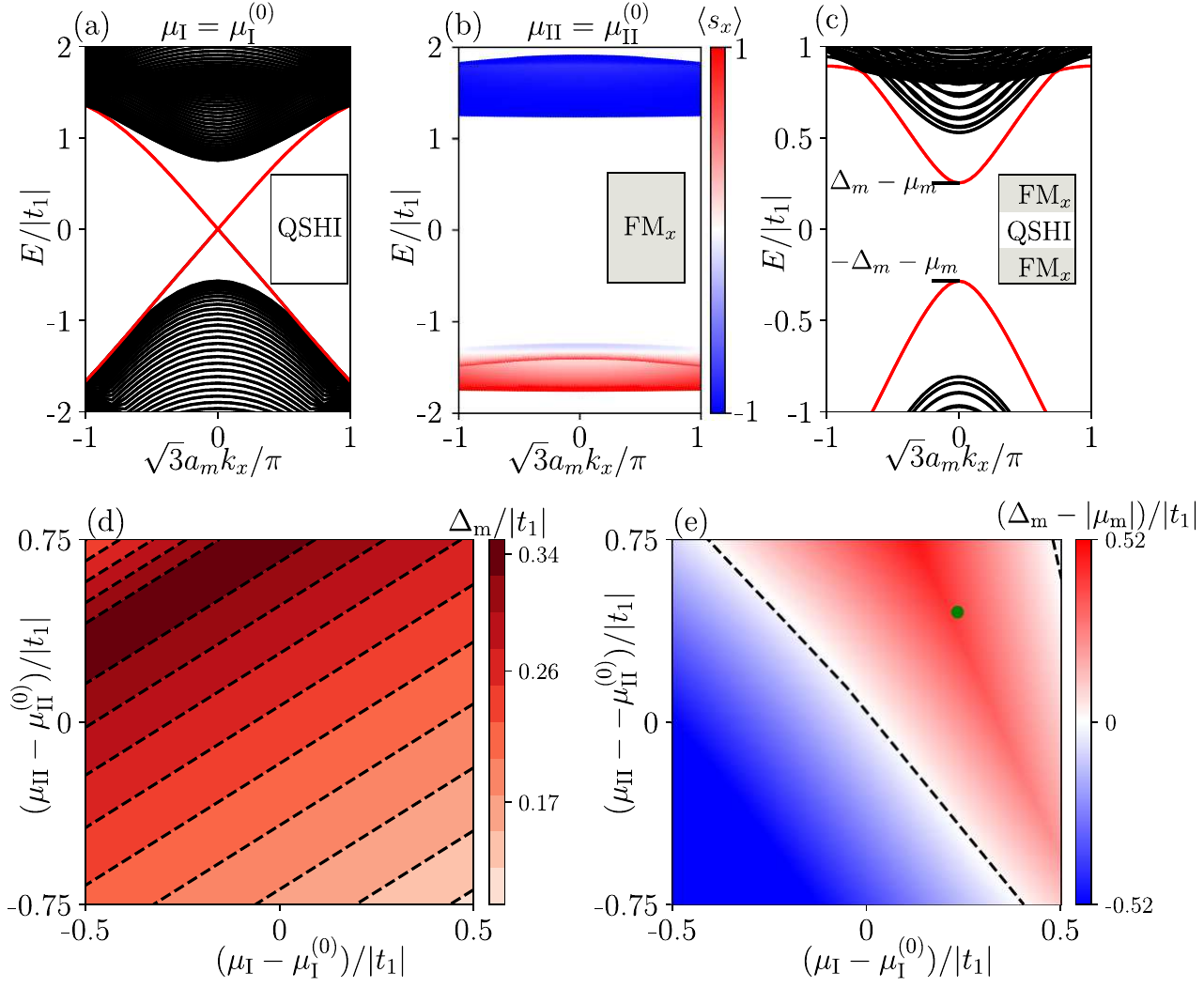}
\caption{(a), (b) Energy spectra of QSHI and FM$_{x}$ states in a cylinder geometry. In (a), the chemical potential $\mu_{\text{I}}^{(0)}$ is taken to set the Dirac point of edge states (red lines) to be at zero energy. In (b), the chemical potential $\mu_{\text{II}}^{(0)}$ is taken to set the middle of the charge gap to be at zero energy. (c) Energy spectra in a cylinder geometry for $\text{FM}_{x}$/QSHI/$\text{FM}_{x}$ junction. $2\Delta_{\text{m}}$ and $\mu_{\text{m}}$ denote the gap and effective chemical potential of the edge states, respectively. (d), (e) Numerical results of $\Delta_{\text{m}}$ and $\Delta_{\text{m}}-|\mu_{\text{m}}|$ as functions of $\mu_{\text{I}}$ and $\mu_{\text{II}}$. In (e), $\Delta_{\text{m}}-|\mu_{\text{m}}|=0$ along the black dashed lines. In (c), parameters $\mu_{\text{I}}$ and $\mu_{\text{II}}$ are fixed at the green point in (e).}
\label{fig2}
\end{figure}

We present the mean-field phase diagrams at  $\nu_h=2$ and $\nu_h=1$ as functions of $\theta$ and $V_z$ at $U=5|t_1|$  in Figs.~\ref{fig1}(b) and \ref{fig1}(c), respectively. Here a phenomenological value of $U$ is chosen to realize interaction-driven phases. At $\nu_h=2$, the system with $V_z=0$ would be in the QSHI phase in the non-interacting limit ($U=0$) but is driven to the AFM$_x$ phase by a finite Hubbard repulsion $U$. Remarkably, the QSHI phase can reemerge in the presence of a finite $V_z$ [Fig.~\ref{fig1}(b)]. In addition, the AFM$_z$ CI phase can also be stabilized by the $V_z$ field. The system is finally turned into a band insulator by a sufficiently large $V_z$. We note that all these four phases were previously reported for the Kane-Mele-Hubbard model \cite{Jiang2018}. At $\nu_h=1$, the system is in the topologically nontrivial CI phase at small $V_z$ but is driven to the topologically trivial magnetic phases (including FM$_z$, FM$_x$, and 120$^{\circ}$ AFM$_{xy}$ states) by large $V_z$. This $V_z$-tuned topological phase transition was recently experimentally observed 
\cite{Anderson2023,Cai20231,Zeng20231,Park20231,Xu20231,Foutty2023}. Similar theoretical phase diagrams at $\nu_h=1$ can be found in Refs.~\cite{Devakul2021,Li2023}.
Since multiple phases can be stabilized by tuning $V_z$ and $\nu_h$ at a fixed twist angle $\theta$,
electric gate-defined junctions formed between different quantum phases can be experimentally realized within one sample. In the following, we study junctions formed between the $\nu_h=2$ QSHI phase and magnetic phases with in-plane magnetization such as $\nu_h=1$ FM$_x$ and $\nu_h=2$ AFM$_x$ phases.

\textit{Magnetic proximity effect}.---
To study the  QSHI/FM$_x$ junction, without loss of generality, we take $\theta=1.5^{\circ}$ and $V_z=4|t_1|$, where the system is in the QSHI phase at $\nu_h=2$ [Fig.~\ref{fig1}(b)] and the $\text{FM}_{x}$ phase at $\nu_h=1$ [Fig.~\ref{fig1}(c)], respectively. We first characterize the two phases separately.
The  QSHI phase can be described by the mean-field Hamiltonian,
\beqn
H_{\text{MF}}^{\text{I}}=H_{\text{KM}}+\sum_{i,\alpha}(\frac{V_z \ell_{\alpha}+U\langle \hat{n}_{i\alpha}\rangle}{2}-\mu_{\text{I}})\hat{n}_{i\alpha},
\label{QH}
\eeqn
where $\langle \hat{n}_{i\alpha}\rangle$ is the mean-field average value of  density operator $\hat{n}_{i\alpha}$ and $\mu_{\text{I}}$ is the chemical potential. In Fig.~\ref{fig2}(a), we present the energy spectra of $H_{\text{MF}}^{\text{I}}$ in a cylinder geometry with periodic boundary condition along the $x$ direction (armchair direction) and open boundary condition along the $y$ direction (zigzag direction). The red bands in Fig.~\ref{fig2}(a) highlight the gapless helical edge states protected by time-reversal symmetry.

Similarly,  the mean-field Hamiltonian for the $\text{FM}_{x}$ phase is,
\beqn
&&H_{\text{MF}}^{\text{II}}=H_{\text{KM}}+\sum_{i,\alpha}(\frac{V_z \ell_{\alpha}+U\langle \hat{n}_{i\alpha}\rangle}{2}-\mu_{\text{II}})\hat{n}_{i\alpha}\nonumber\\
&&\quad\quad\quad-\frac{U}{2}\sum_{i,\alpha}\langle \hat{S}_{i\alpha}^{x}\rangle\hat{S}_{i\alpha}^{x},
\label{MH}
\eeqn
where  $ \hat{ S}_{i\alpha}^{x}=\sum_{ss^{\prime}}c_{i\alpha s}^{\dagger} (s_{x})_{ss^{\prime}}c_{i\alpha s^{\prime}}$ with $s_x$ being the Pauli matrix in the spin space and $\mu_{\text{II}}$ is the chemical potential. In Fig.~\ref{fig2}(b), we plot the energy spectra of $H_{\text{MF}}^{\text{II}}$ in a cylinder geometry, which shows a sizable charge gap at $\nu_h=1$.  The color of bands encodes the in-plane spin expectation value $\langle s_x\rangle$. The bands 
above the gap
have larger $|\langle s_x\rangle|$ than that of bands right below the gap, because the former (latter) states mainly reside in A (B) sites, and magnetic moments mainly develop at A sites for $V_z>0$. The interaction-driven magnetic term can be approximated as $-U\langle \hat{S}_{i\alpha}^{x}\rangle (\sigma_0+\sigma_z)s_x/4$ with Pauli matrix $\sigma_z$ acting in sublattice space.

\begin{figure}
\centering
\includegraphics[width=3.5in]{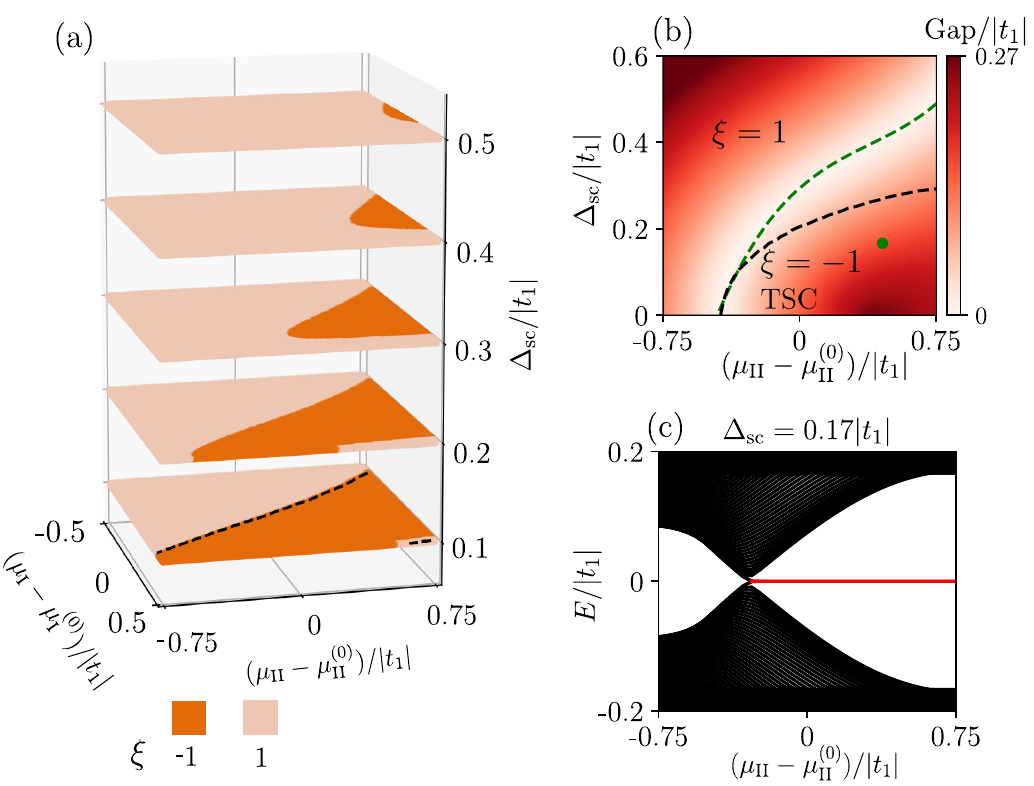}
\caption{Results for the the SC-QSHI/FM$_x$ junction. (a) The phase diagram characterized by the $Z_2$ topological invariant $\xi$ in a cylinder geometry as a function of $\mu_{\text{I}}$, $\mu_{\text{II}}$, and $\Delta_{\text{sc}}$.  (b) The energy gap in a cylinder geometry as a function of $\mu_{\text{II}}$ and  $\Delta_{\text{sc}}$ at a fixed $\mu_{\text{I}}$. The green dashed line marks the topological phase transition and the black dashed line is obtained through the effective edge theory of Eq.~\eqref{Hedge}. (c) The energy spectra under open boundary conditions in both directions with respect to $\mu_{\text{II}}$ at fixed $\mu_{\text{I}}$ and $\Delta_{\text{sc}}$. The red bands label the MZMs. In (b) and (c), $\mu_{\text{I}}$ takes the same value as that in Fig.~\ref{fig2}(c).}
\label{fig3}
\end{figure}

We now study a symmetric $\text{FM}_{x}$/QSHI/$\text{FM}_{x}$ junction with the junction interface along the armchair direction. We take the chemical potentials $\mu_{\text{I}}$ and $\mu_{\text{II}}$ for the two phases  as independently tunable parameters and choose them such that zero energy is always within the bulk charge gap of each phase. The energy spectra of the junction in the cylinder geometry are  shown in Fig.~\ref{fig2}(c). The edge states of QSHI  become gapped by the magnetic proximity effect. From the spectrum, we define $2\Delta_{\text{m}}$ as the gap and $\mu_m$ as the effective chemical potential of the edge states.

To have a qualitative understanding,
we consider the spinor part of the armchair edge states $\Psi_{1,2}(y)$ of the QSHI for the semi-infinite system with $y>0$, which generally takes the form \cite{Pan2019}
\beqn
&&\chi_1=\left|\sigma_y=1\right\rangle \otimes|s_z=-1\rangle, \nonumber\\
&& \chi_2=\left|\sigma_y=-1\right\rangle\otimes|s_z=1\rangle.
\label{et}
\eeqn
Therefore, the two edge states $\Psi_{1,2}$ can be coupled by the sublattice-dependent magnetic term $-U\langle \hat{S}_{i\alpha}^{x}\rangle (\sigma_0+\sigma_z)s_x/4$, but not by a sublattice independent term \cite{Pan2019,Ren2020}. We note that an out-of-plane magnetic term can not gap out the edge states, which is the reason to use magnetic phases with in-plane magnetization in the junction.

We present the numerical value of $\Delta_{\text{m}}$ as a function of $\mu_{\text{I}}$ and $\mu_{\text{II}}$ in Fig.~\ref{fig2}(d).  
Here $\Delta_{\text{m}}$ depends only on the difference $\mu_{\text{II}}-\mu_{\text{I}}$.  We find that $\Delta_{\text{m}}$ becomes larger (smaller) as the QSHI edge state is tuned in energy closer to the conduction (valence) bands of the FM$_x$ phase, which is consistent with the band-dependent magnetization shown in Fig.~\ref{fig2}(b).
In Fig.~\ref{fig2}(e), we present the numerical value of $\Delta_m-|\mu_m|$ as a function of $\mu_{\text{I}}$ and $\mu_{\text{II}}$.
When $\Delta_m>|\mu_m|$, there are no states at the Fermi level and the system behaves as an insulator, which is important to realize MZMs as we will explain; otherwise, the system behaves as a metal.

\textit{MZMs in SC-QSHI/FM$_x$ junction}.--- 
We turn to study SC-QSHI/FM$_x$ junction, where the QSHI region has proximitized superconducting pairing potential $\Delta_{\text{sc}}\sum_{i \alpha}c_{i\alpha\uparrow}^{\dagger}c_{i\alpha\downarrow}^{\dagger}+h.c.$.
This junction  can be viewed as a 1D system with translational symmetry along the interface direction, which belongs to the D symmetry class of
the Atland-Zirnbauer classification and can be characterized by the $Z_2$ topological invariant $\xi$ if its energy spectra are fully gapped. Here $\xi$ can be defined by the Pfaffian value of the Hamiltonian of the system in the Majorana basis \cite{Kitaev2001}, and $\xi=-1$ (1) corresponds to a topological (trivial) phase with (without) robust MZMs at the boundary. 

We present the topological phase diagrams of the junction in a cylinder geometry as a function of $\mu_{\text{I}}$, $\mu_{\text{II}}$, and $\Delta_{\text{sc}}$ in Fig.~\ref{fig3}(a). The phase diagrams can be understood as follows. In the parameter space where $\Delta_m>|\mu_m|$ (enclosed by the black dashed lines), an infinitesimal (but nonzero) pairing potential $\Delta_{\text{sc}}$ can drive this system into a TSC. This can be understood by examining a system with open boundary conditions in both directions, where the armchair
and zigzag edges of QSHI have the magnetic-dominated gap and superconducting-dominated gap, respectively.  In this case, there are isolated MZMs \cite{FuKane2009} at the ends of the interface in the SC-QSHI/$\text{FM}_{x}$ junction [Fig.~\ref{fig1}(e)].  The TSC region becomes narrower with increasing $\Delta_{\text{sc}}$ [Fig.~\ref{fig3}(a)], which can be attributed to the gap closing at the junction interface. This topological phase transition can be revealed by the low-energy effective Hamiltonian of the armchair edge states 
\beqn
\tilde{\mathcal{H}}=k_x\tau_0s_z+\Delta_m\tau_zs_x-\mu_m\tau_zs_0+\tilde{\Delta}_{\text{sc}}\tau_ys_y,
\label{Hedge}
\eeqn
where the Pauli matrices $\tau_{y,z}$ act on the particle-hole (Nambu) space and $\tilde{\Delta}_{\text{sc}}$ is the effective pairing potential for the edge states. The energy spectrum of $\tilde{\mathcal{H}}$ is closed when $\tilde{\Delta}_{\text{sc}}=\sqrt{\Delta_m^2-\mu_m^2}$, which is associated with a topological phase transition.

\begin{figure}
\centering
\includegraphics[width=3.4in]{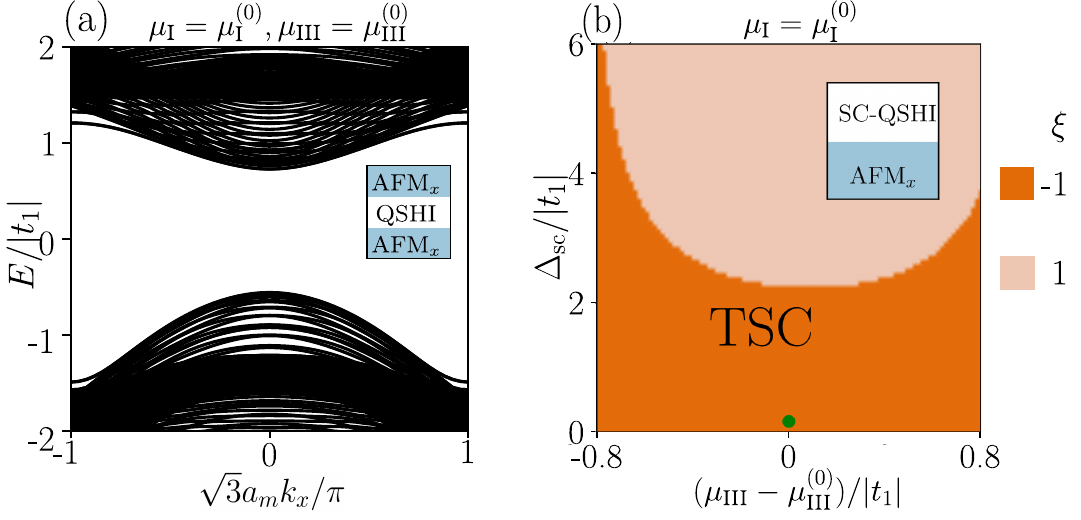}
\caption{(a) The energy spectra of the AFM$_x$/QSHI/AFM$_x$ junction (inset plots) in a cylinder geometry. (b) Topological phase diagram as a function of $\mu_{\text{III}}$  and $\Delta_{\text{sc}}$ for the SC-QSHI/AFM$_x$ junction in a cylinder geometry. $\mu_{\text{III}}^{(0)}$ is taken to set the middle of the charge gap of the AFM$_x$ state to be at zero energy. In (a) and (b), we use $\theta=1.5^{\circ}$ and $V_z=2.5|t_1|$ for the AFM$_x$ state.}
\label{fig4}
\end{figure}

We present the numerical value of the energy gap of the junction in a cylinder geometry as a function of $\Delta_{\text{sc}}$ and $\mu_{\text{II}}$ at a fixed $\mu_{\text{I}}$ in Fig.~\ref{fig3}(b). The green dashed line characterizes the topological phase transition associated with the gap closing. The black dashed line plots $\tilde{\Delta}_{\text{sc}}=\sqrt{\Delta_m^2-\mu_m^2}$. The deviation between the two lines implies that the effective pairing potential $\tilde{\Delta}_{\text{sc}}$ is smaller than ${\Delta}_{\text{sc}}$. This is because the magnetic proximity effect suppresses the superconducting pairing for the edge states. To further elaborate on the topological phase transition, we present the open-boundary energy spectra as a function of $\mu_{\text{II}}$ at fixed $\mu_{\text{I}}$ and $\Delta_{\text{sc}}$ in Fig.~\ref{fig3}(c), where the red bands represent the MZMs.

\textit{MZMs in SC-QSHI/AFM$_x$ junction}.---
We now show that MZMs can also be realized by constructing the SC-QSHI/AFM$_x$ junction. The $\nu_h=2$ AFM$_{x}$ state can also be described by the mean-field Hamiltonian in Eq.~\eqref{MH} but with $\langle \hat{S}_{i\text{A}}^{x}\rangle=-\langle \hat{S}_{i\text{B}}^{x}\rangle$, leading to the magnetic term $-U\langle \hat{S}_{iA}^{x}\rangle\sigma_zs_x/2$. This term can gap the armchair edge states of QSHI described by Eq.~\eqref{et}.  In Fig.~\ref{fig4}(a), we plot the energy spectra of the AFM$_x$/QSHI/AFM$_x$ junction in a cylinder geometry, which are fully gapped as expected.

When adding the superconducting pairing potential to the QSHI phase, we obtain the SC-QSHI/AFM$_x$ junction, as shown in the inset of Fig.~\ref{fig4}(b). Similar to the SC-QSHI/FM$_x$ junction, there are two robust MZMs at the ends of the interface of the SC-QSHI/AFM$_x$ junction [Fig.~\ref{fig1}(f)]. In Fig.~\ref{fig4}(b), we present the phase diagram as a function of $\mu_{\text{III}}$ (i.e., the chemical potential
 in AFM$_x$ region) and $\Delta_{\text{sc}}$ at  fixed  $\mu_{\text{I}}$.  We find that the TSC phase is relatively robust  with respect to the variations of $\mu_{\text{III}}$ and $\Delta_{\text{sc}}$ [ Fig.~\ref{fig4}(b)]. This stability can be traced to the large magnetic energy gap opened by the magnetic proximity effect in the QSHI/AFM$_x$ junction.

\textit{Discussion}.---
In summary, we introduce twisted TMD homobilayers as a potential new platform to realize MZMs. The key of our proposal is to achieve gate-defined QSHI/FM$_x$ and QSHI/AFM$_x$ junctions, which is feasible because of the gate-tunable phase diagram and the recent experimental advances in nanodevice fabrication \cite{Díez-Mérida2023}. When bringing in the superconducting proximity effect for the QSHI state, the system can be topologically nontrivial and host MZMs, which can be detected, for example, by tunneling experiments through zero-bias conductance peaks \cite{Liu2012}.
Topological phase transitions can be readily tuned electrically without involving any external magnetic field, which is a potential advantage of our scheme. 

The MZMs are protected by particle-hole symmetry and can not be removed as long as the energy gap of the system persists.  We use junctions with the interface along the armchair direction for illustration, but we find that MZMs can be realized for any generic interface direction \cite{supp}.  In the calculation, the superconducting pairing potential is assumed to be uniform across the QSHI region and identical for the A and B sublattices.  This assumption can be relaxed, since the main physics is governed by the edge states \cite{supp}. While we demonstrate that the TSC phase diagram is electrically tunable by adjusting the chemical potentials within the bulk charge gap, the parameter $V_z$ can also serve as a tuning knob.    In the search for MZMs, disorder effects often can not be ignored  \cite{Pan2020c}. The effect of disorder and interaction physics beyond mean-field theory in our scheme requires future study.  

Fractional quantum anomalous Hall effect was recently observed in $t$MoTe$_2$ \cite{Cai20231,Zeng20231,Park20231,Xu20231}. In contrast to Landau levels formed in an external magnetic field, twisted TMD homobilayers respect time-reversal symmetry, unless it is spontaneously broken, for example, in the (fractional) quantum anomalous Hall states. As a theoretical possibility, time-reversal symmetric fractional quantum spin Hall insulators \cite{Levin2009,Qi2011} could also emerge in twisted TMD homobilayers. In the junctions studied in this work, parafermionic zero modes \cite{Lindner2012,Clarke2013} can be realized when the QSHI state is replaced by its fractionalized cousins (i.e., the fractional quantum spin Hall insulators).

We thank Haining Pan, Chun-Xiao Liu, and Jie Wang for helpful discussion. This work is supported by National Natural Science Foundation of China (Grant No. 12274333), National Key Research and Development Program of China (Grants No. 2022YFA1402401 and No. 2021YFA1401300), and start-up funding of Wuhan University. We also acknowledge support by Key Research and Development Program of Hubei Province (Grant No. 2022BAA017).

\bibliography{reference}

\clearpage

\begin{widetext}
\begin{center}
\begin{large}
\textbf{Supplemental Material for ‘‘Majorana zero modes in twisted transition metal dichalcogenide homobilayers"}
\end{large}
\end{center}

This Supplemental Material includes the following four sections:
(1) Continuum model and tight-binding model for moir\'e bands;
(2) Hartree-Fock mean-field calculation;
(3) Phase diagrams of junctions with the interface along a generic direction;
(4) Phase diagram with different superconducting pairing amplitudes on A and B sublattices.

\section{Continuum model and tight-binding model for moir\'e bands}
The low-energy continuum Hamiltonian for twisted TMD homobilayers at $+K$ valley is \cite{Wu2019},
\beqn
H_{+K}=\begin{pmatrix}
-\frac{\hbar^2 (\bm k-\boldsymbol{\kappa}_+)^2}{2 m^*} +\Delta_{+}(\bm r) & \Delta_{\text{T}}(\bm r) \\
\Delta_{\text{T}}^{\dagger}(\bm r) & -\frac{\hbar^2 (\bm k-\boldsymbol{\kappa}_-)^2}{2 m^*} +\Delta_{-}(\bm r)
\end{pmatrix},\nonumber\\
\label{mh}
\eeqn
where  $\bm\kappa_{\pm}=4\pi/3a_m(-\sqrt{3}/2,\mp 1/2)$  are located at the corners of the moiré Brillouin zone, the off-diagonal terms describe the interlayer tunneling $\Delta_{\text{T}}(\bm r) $, and the diagonal terms describe the momentum-shifted kinetic energy with the effective mass $m^{*}=0.43m_e$ ($m_e$ is
the rest electron mass), plus the intralayer potential $\Delta_{+/-}(\bm r)$. The intralayer potential and interlayer tunnelings are parametrized as follows,
\beqn
\Delta_{\pm} (\bm r)= 2 V \sum_{j=1, 3, 5} \cos(\bm g_j \cdot \bm r \pm \psi),\nonumber\\
\Delta_{\text{T}}(\bm r) = w (1+ e^{-i \bm g_2 \cdot \bm r}+ e^{-i \bm g_3 \cdot \bm r}),
\label{Potential}
\eeqn
where $V$ and $\psi$, respectively, characterize the amplitude
and spatial pattern of the moiré potential, $\bm g_j$ is the reciprocal lattice vector obtained by counterclockwise
rotation of $\bm g_1=4\pi/\sqrt{3}a_m(1,0)$ with angle $(j-1)\pi/3$, and $w$ is the interlayer tunneling strength.
The model parameters $(V, \psi, w)$ can be obtained by fitting to the first principles band structures. For twisted WSe$_2$, we have $(V, \psi, w)\approx $ (9 meV, $128^{\circ}$, 18 meV) \cite{Devakul2021}.

\begin{figure*}
\centering
\includegraphics[width=4.7in]{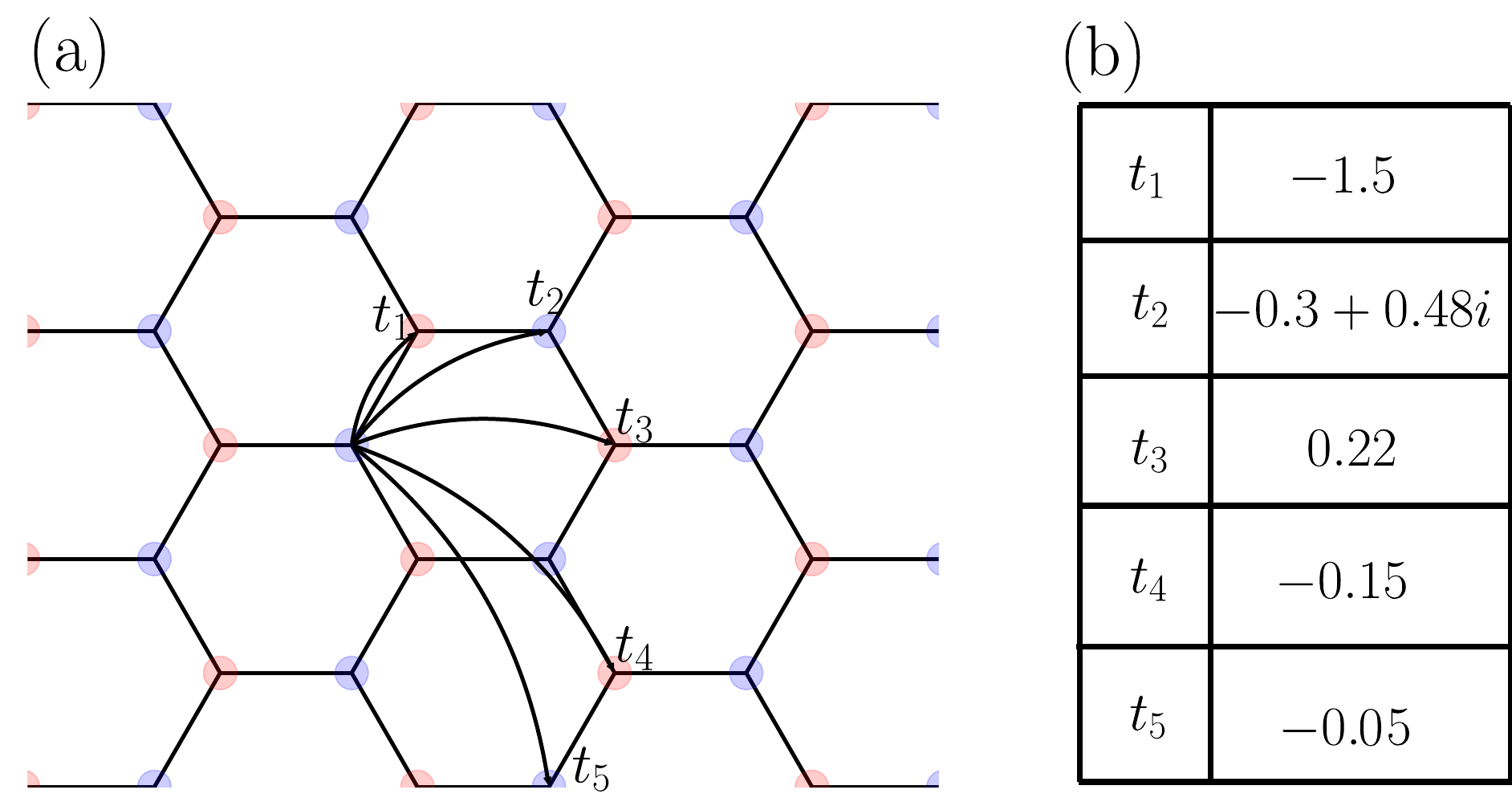}
\caption{(a) Schematical illustration of the hopping parameters $t_{1,2,3,4,5}$ in the generalized Kane-Mele model. (b) Numerical values of  $t_{1,2,3,4,5}$  (in the unit of meV) at twist angle $\theta=1.5^{\circ}$ for Hamiltonian in the spin-up sector.}
\label{fig5}
\end{figure*}

The moir\'e band topology can be tuned by the twist angle $\theta$. For $\theta<\theta_1$ ($\theta_1\approx 1.5^{\circ}$),  the topmost two bands carry Chern number 1 and $-1$, respectively. In this case, the tight-binding model for the topmost two bands can be explicitly derived by constructing the Wannier states, which turns out to be the generalized Haldane model (Kane-Mele model if two valleys are included) on a honeycomb lattice \cite{Wu2019}. The two Wannier states localized at A and B sublattices are mainly distributed at the bottom and top layers, respectively. Thus,  a vertical electric field can generate a staggered sublattice potential.Moreover, the hopping parameters for the tight-binding model can also be calculated based on the constructed Wannier states \cite{Devakul2021}. 
In Fig.~\ref{fig5}(a), we present a schematic illustration of the hopping parameters up to the fifth order and present their numerical value at $\theta=1.5^{\circ}$ in Fig.~\ref{fig5}(b).

\begin{figure*}
\centering
\includegraphics[width=7in]{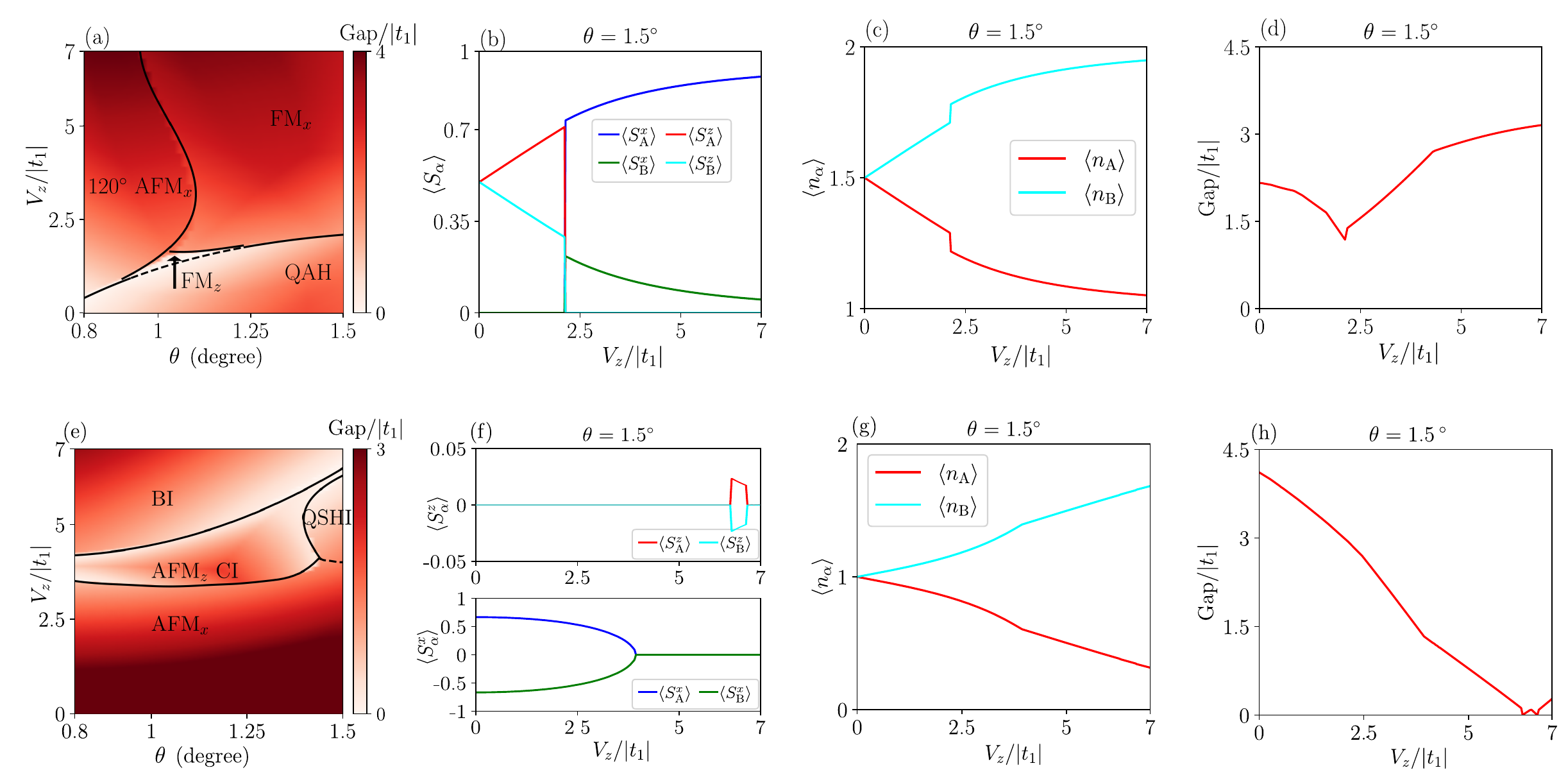}
\caption{(a), (e) Bulk energy gap and quantum phase diagrams at $\nu_h=1$ and $\nu_h=2$ as functions of $\theta$ and $V_z$ by fixing $U=5|t_1|$. The solid and dashed lines mark the first-order and continuous phase transitions, respectively.  (b)-(d) and (e)-(h) The spin magnetic moments, sublattice-dependent occupation number, and bulk energy gap as functions of $V_z$ by fixing $\theta=1.5^{\circ}$ at  $\nu_h=1$ and $\nu_h=2$, respectively. }
\label{fig6}
\end{figure*}

\section{Hartree-Fock mean-field calculation}
The Hamiltonian of the generalized Kane-Mele Hubbard model can be written as
\beqn
&&H=H_{\text{KM}}+U\sum_{i,\alpha}\hat{n}_{i\alpha\uparrow}\hat{n}_{i\alpha\downarrow}+\frac{V_z}{2}\sum_{i,\alpha}\ell_{\alpha}\hat{n}_{i\alpha },\nonumber\\
&&H_{\text{KM}}= t_1 \sum_{\langle ij\rangle,\alpha\neq \beta,s} c^\dagger_{i\alpha s} c_{j\beta s} +|t_2| \sum_{\langle \langle ij\rangle\rangle,\alpha,s} e^{i\phi s \epsilon_{ij}} c^\dagger_{i\alpha s} c_{j\alpha s}+\cdots,
\label{Ham}
\eeqn
where subscripts $\alpha$ and $\beta$ denote the A or B sublattice, $s$ is the spin index, $U$ represents the onsite Coulomb repulsion, and $V_z$ denotes the staggered sublattice potential strength. Within the Hartree-Fock approximation, the onsite Coulomb interaction can be decomposed as
\beqn
U\sum_{i,\alpha}\hat{n}_{i\alpha\uparrow}\hat{n}_{i\alpha\downarrow} \approx \frac{U}{2}\sum_{i,\alpha}(\langle \hat{n}_{i\alpha}\rangle\hat{n}_{i\alpha}-\langle \hat{n}_{i\alpha }\rangle^2/2)-\nonumber\\
\quad\quad\quad\quad\frac{U}{2}\sum_{i,\alpha,\rho=x,y,z}(\langle \hat{S}_{i\alpha}^{\rho}\rangle\hat{S}_{i\alpha}^{\rho}-\langle \hat{S}_{i\alpha}^{\rho}\rangle^2/2),
\eeqn
where $ \hat{ S}_{i\alpha}^{\rho}=\sum_{ss^{\prime}}c_{i\alpha s}^{\dagger} (s_{\rho})_{ss^{\prime}}c_{i\alpha s^{\prime}}$ with $ s_{x,y,z}$ being the Pauli matrices acting on the spin space. Then the full mean-field Hamiltonian is 
\beqn
&&H=H_{\text{KM}}+\frac{V_z}{2}\sum_{i,\alpha}\ell_{\alpha}\hat{n}_{i\alpha}+\frac{U}{2}\sum_{i,\alpha}\langle \hat{n}_{i\alpha}\rangle\hat{n}_{i\alpha}-\frac{U}{2}\sum_{i,\alpha,\rho}\langle \hat{S}_{i\alpha}^{\rho}\rangle\hat{S}_{i\alpha}^{\rho}.
\eeqn
where the constant terms are dropped. The values of $\langle \hat{n}_{i\alpha}\rangle$ and $\langle \hat{S}_{i\alpha}^{\rho}\rangle$ can be obtained by a self-consistent calculation.

\begin{figure}
\centering
\includegraphics[width=6.3in]{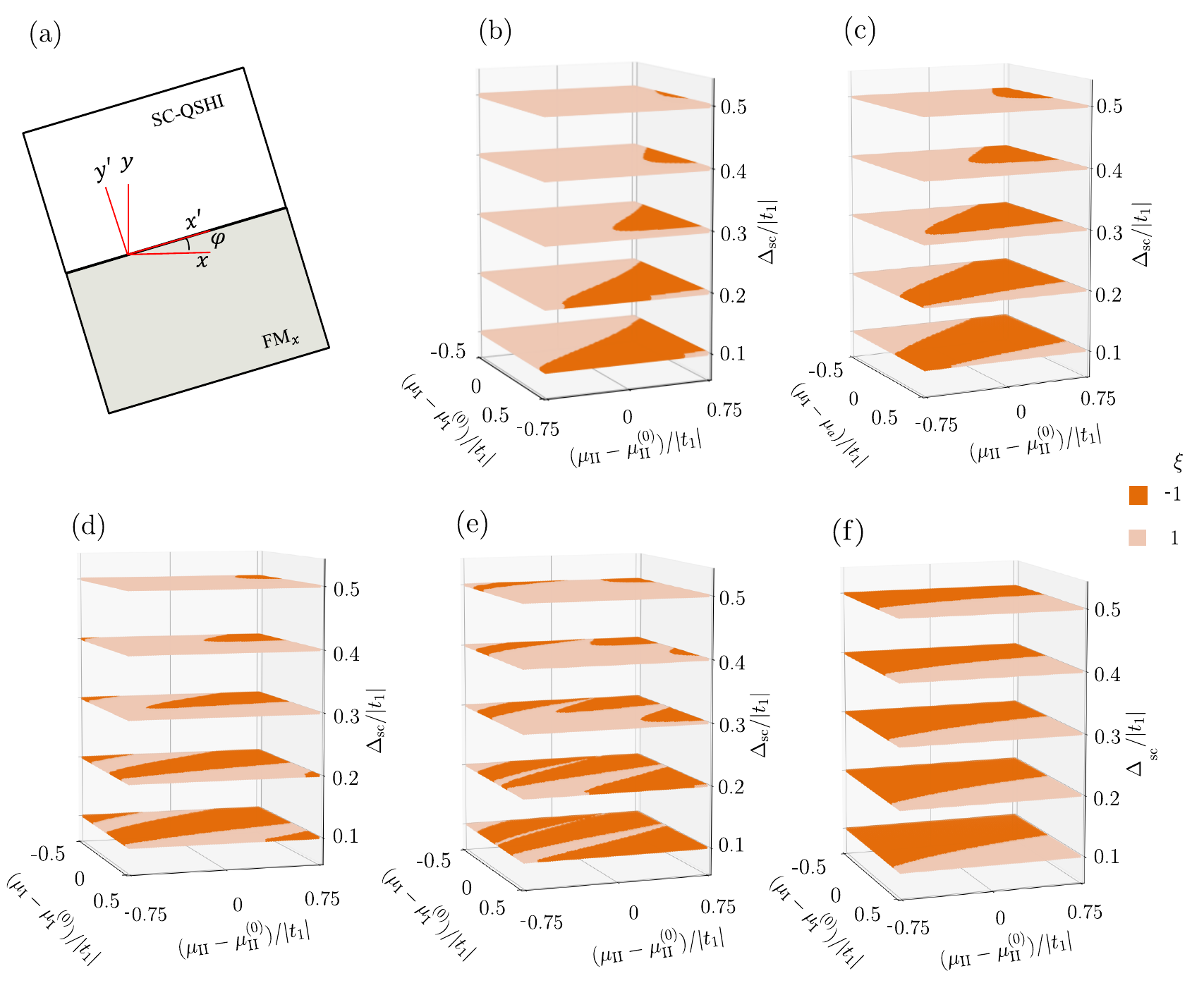}
\caption{ (a) Schematic plot of the SC-QSHI/FM$_x$ junction with the interface along a generic direction. (b)-(f) Topological phase diagram of the SC-QSHI/FM$_x$ junction with the interface along the direction defined by $\vec{e}_{x^{\prime}}=m\vec{e}_1+n\vec{e}_2$ with $(m,n)=(1,1), (1,2), (1,3), (1,4)$, and $(0,1)$, respectively. The corresponding angle between $\vec{e}_{x^{\prime}}$ and $\bm e_x$ is $0^{\circ}$, $2.12^{\circ}$, $4.76^{\circ}$, $6.84^{\circ}$, and $30^{\circ}$, respectively. }
\label{fig7}
\end{figure}

\begin{figure}
\centering
\includegraphics[width=6.3in]{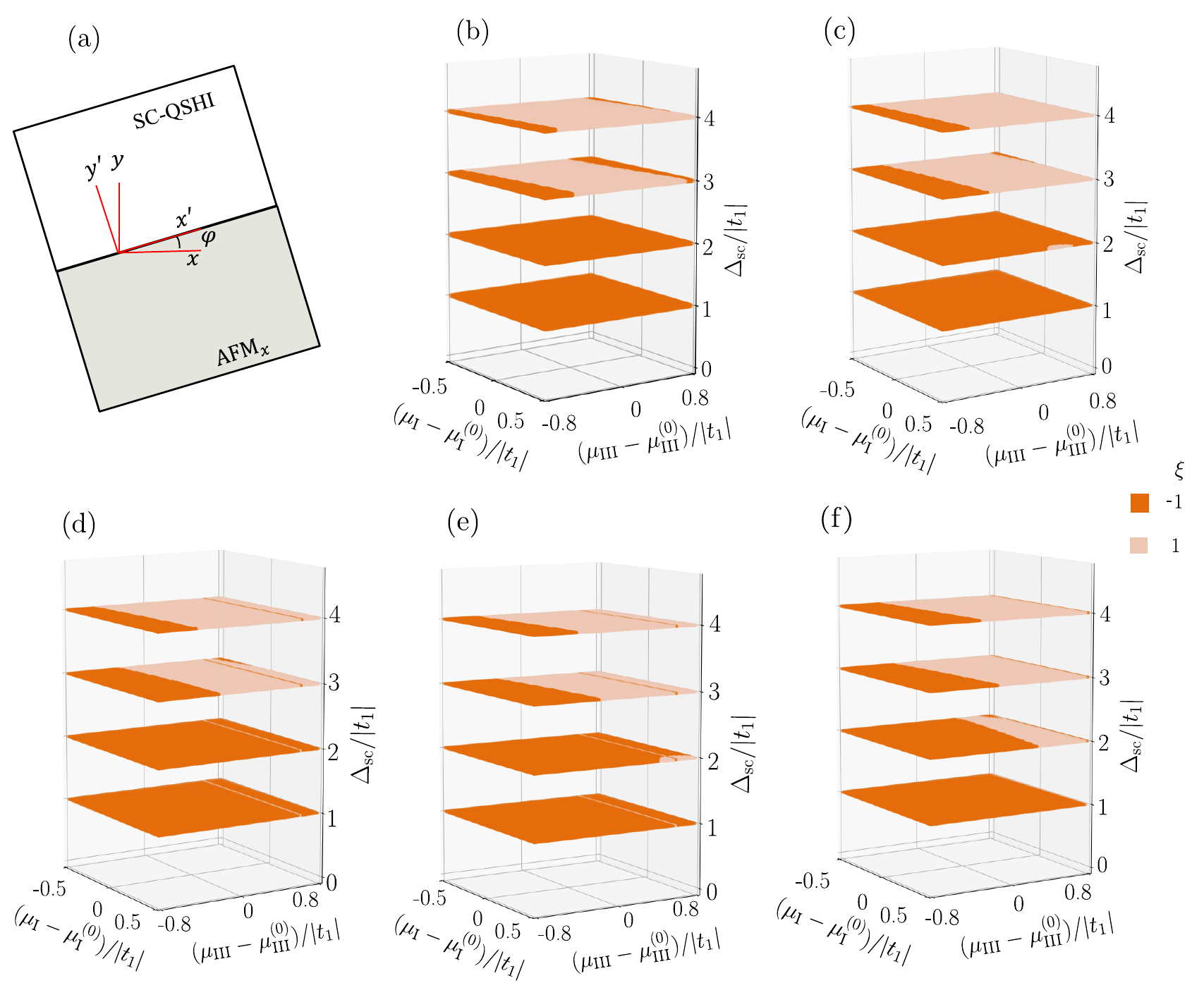}
\caption{ (a) Schematic plot of the SC-QSHI/AFM$_x$ junction with the interface along a generic direction. (b)-(f) Topological phase diagram of the SC-QSHI/AFM$_x$ junction with the interface along the direction defined by $\vec{e}_{x^{\prime}}=m\vec{e}_1+n\vec{e}_2$ with $(m,n)=(1,1), (1,2), (1,3), (1,4)$, and $(0,1)$, respectively. The corresponding angle between $\vec{e}_{x^{\prime}}$ and $\bm e_x$ is $0^{\circ}$, $2.12^{\circ}$, $4.76^{\circ}$, $6.84^{\circ}$, and $30^{\circ}$, respectively.}
\label{fig9}
\end{figure}

We present the bulk energy gap at $\nu_h=2$ and $\nu_h=1$ as a function of $\theta$ and $V_z$ at $U=5|t_1|$ in Figs.~\ref{fig6} (a) and \ref{fig6} (e), respectively. At $\nu_h=2$, the phase diagram includes the quantum spin Hall insulator (QSHI), the in-plane antiferromagnetic state (AFM$_x$), the out-of-plane antiferromagnetic Chern insulator (AFM$_z$ CI) with a Chern number $C$ of 1, and the band insulator (BI). At $\nu_h=1$, the phase diagram hosts the spin-polarized Chern insulator (CI) with $C=1$, the topologically trivial out-of-plane ferromagnetic state (FM$_z$), the in-plane ferromagnetic state (FM$_x$), and the 120$^{\circ}$ antiferromagnetic state (120$^{\circ}$ AFM$_{xy}$). We note that all these phases at $\nu_h=2$ and $\nu_h=1$ have been previously studied for the Kane-Mele-Hubbard model  \cite{Jiang2018,Devakul2021}. In Figs.~\ref{fig6}(b)-\ref{fig6}(d) and Figs.~\ref{fig6}(f)-\ref{fig6}(h), we present the numerical values of $\langle \hat{S}_{\alpha}\rangle$, $\langle \hat{n}_{\alpha}\rangle$, and bulk energy gap at $\nu_h=2$ and $\nu_h=1$, respectively, as functions of $V_z$ by fixing $\theta=1.5^{\circ}$. In the main text, we take $V_z=4|t_1|$ to obtain the QSHI state at $\nu_h=2$ and  the FM$_x$ state at $\nu_h=1$, which leads to ($\langle \hat{n}_\text{A}\rangle, \langle \hat{n}_{\text{B}}\rangle$)=(0.4,1.6) in the QSHI state and ($\langle \hat{n}_\text{A}\rangle, \langle \hat{n}_{\text{B}}\rangle,\langle \hat{S}_\text{A}^{x}\rangle,\langle \hat{S}_\text{B}^{x}\rangle)=(1.11,1.89,0.84,0.11)$ in the FM$_x$ state.
We take $V_z=2.5|t_1|$ to obtain the AFM$_x$ state at $\nu_h=2$, which gives rise to ($\langle \hat{n}_\text{A}\rangle, \langle \hat{n}_{\text{B}}\rangle,\langle S_\text{A}^{x}\rangle,\langle S_\text{B}^{x}\rangle)=(0.82,1.18,0.53,-0.53)$.
We note that the hole filling factor $\nu_h$ is related to the occupation number as $\nu_h=4-(\langle \hat{n}_\text{A}\rangle+\langle \hat{n}_{\text{B}}\rangle)$.

\section{Phase diagram of junctions with the interface along a generic direction}
In the main text, we demonstrate that MZMs can be realized in the SC-QSHI/FM$_x$ junction with the interface
along the armchair direction. Here we show that MZMs can be realized for a generic interface direction.

We first define the  primitive lattice vectors for the honeycomb lattice as 
$\vec{e}_1=a_m(\sqrt{3}/2,-1/2)$ and $\vec{e}_2=a_m(\sqrt{3}/2,1/2)$. Then the armchair ($x$) and zigzag $(y)$ directions are defined by $\vec{e}_x=\vec{e}_1+\vec{e}_2$ and $\vec{e}_y=-\vec{e}_1+\vec{e}_2$, respectively. 
In Fig.~\ref{fig7}(a), we present a schematic illustration of the SC-QSHI/FM$_x$ junction with the interface along an arbitrary $x^{\prime}$ direction defined by a vector $\vec{e}_{x^{\prime}}$.
We note that only when $\vec{e}_{x^{\prime}}$ is parallel to $m\vec{e}_1+n\vec{e}_2$ with $m$ and $n$ being integers, the system have the lattice translational symmetry along the $x^{\prime}$ direction.  Without loss of generality, we focus on such interface directions. 

In Figs.~\ref{fig7}(b)-\ref{fig7}(f), we present the topological phase diagram as a function of $\mu_{\text{I}}$, $\mu_{\text{II}}$ and $\Delta_{\text{sc}}$ for the  SC-QSHI/FM$_x$ junction along the interface directions defined, respectively, by $(m,n)=(1,1), (1,2), (1,3), (1,4)$, and $(0,1)$. We find that the SC-QSHI/FM$_x$ junction with the interface along all these five directions can host TSC. Similarly, TSC can also be realized in the SC-QSHI/AFM$_x$ junction with the interface along a generic direction, as  illustrated in Fig.~\ref{fig9}.

\begin{figure}
\centering
\includegraphics[width=3.3in]{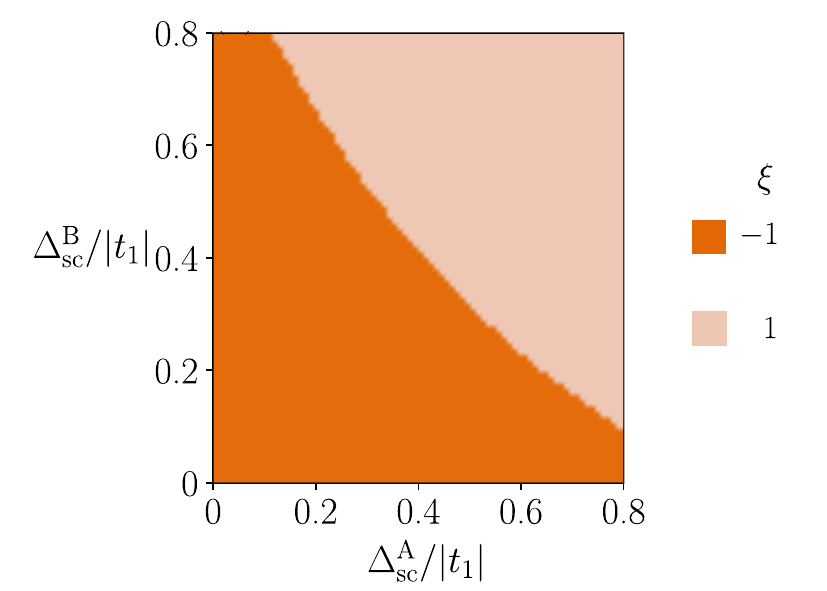}
\caption{ The topological superconductor phase diagram as a function of the superconducting pairing amplitudes $\Delta_{\text{sc}}^{\text{A}}$ and $\Delta_{\text{sc}}^{\text{B}}$ for the A and B sublattices. Other model parameters are the same as those used in Fig. 1(e) of the main text.}
\label{s5}
\end{figure}

\section{Phase diagram with different pairing amplitudes on A and B sublattices}
In the Kane-Mele model simulated by the TMD homobilayers, the Wannier orbitals localized at A and B sublattices are polarized to opposite layers. Thus, when one of the TMD layers is closer to the superconductor, the pairing amplitudes for the A and B sublattices can have different magnitudes.

We present the topological superconductor phase diagram as a function of the superconducting pairing amplitudes $\Delta_{\text{sc}}^{\text{A}}$ and $\Delta_{\text{sc}}^{\text{B}}$ for the A and B sublattices, as shown in Fig.~\ref{s5}. The topological superconductor phase is relatively robust against the difference between $\Delta_{\text{sc}}^{\text{A}}$ and $\Delta_{\text{sc}}^{\text{B}}$.

\end{widetext}

\end{document}